\documentclass{jaa}
\usepackage{graphicx}
\usepackage{hyperref}
\hypersetup{
    colorlinks=true,
    linkcolor=blue,
    filecolor=magenta,      
    urlcolor=cyan,
}
 
\begin{document}\sloppy

%%paper title
%%For line breaks \\ can be used within title
\title{Installation of Solar Chromospheric Telescope at the Indian Astronomical Observatory, Merak}

%%author names are separated by comma (,)
%%use \and before the last author name
%%use a * along with the number separated by comma
%% for the  author for correspondence
%%\textsuperscript{number} is used for affiliation
%%\affilOne, \affilTwo etc., upto \affilTwentyfive is possible
%%Please note the first letter after \affil is capitalised in the command
%%

\author{Ravindra B\textsuperscript{1}, Prabhu Kesavan, Thulasidharen, K. C., Rajalingam, M.,  Sagayanathan, K., Kamath, P. U., Namgyal Dorjey, Angchuk Dorjee, Kemkar, P. M. M., Tsewang Dorjai, Ravinder K. Banyal}
\affilOne{\textsuperscript{1} Indian Institute of Astrophysics, II Block, Koramangala, Bengaluru-560 034, INDIA}

%%escape two column mode for title, affiliation and abstract
%%by giving \twocolumn command as shown

\twocolumn[{

\maketitle

%%include \corres to print the corresponding author Email id
\corres{ravindra@iiap.res.in}

%%include \msinfo for
%%manuscript information such as
%%received, revised and accepted dates
%%
\msinfo{1 June 2018}{1 Sep 2018}

%%abstract
\begin{abstract}
We report the observations of the solar chromosphere from a newly commissioned solar telescope
at the incursion site near Pangong Tso lake in Merak (Leh/Ladakh). This new H$_{\alpha}$ telescope at the Merak site is identical to the Kodaikanal H$_{\alpha}$ telescope. The telescope is installed in the month of August, 2017 at the Merak site. A 20-cm doublet lens with additional re-imaging optics makes the telescope. A Lyot filter with 0.5~\AA~passband isolates the Balmer line of the hydrogen spectra to make the observations of the solar chromosphere. The observations made in H$_{\alpha}$ wavelength delineates the magnetic field directions at the sunspot and the quiet regions. A CCD detector records the images of the chromosphere with a pixel resolution of 0.27$^{\prime\prime}$ and covers  9.2$^{\prime}$ field-of-view. This telescope has a good guiding system that keeps the FOV in the intended position.  We report  the development of control software for tuning the filter unit, control detector system, observations, and calibration of the data to make it useful for the scientific community. Some preliminary results obtained from the Merak H$_{\alpha}$ telescope are also presented. This high altitude facility is a timely addition to regularly available H$_{\alpha}$ images around the globe.
\end{abstract}

%%insert keywords separated by 3 hyphens using \keywords{words}

}]
%%close the twocolumn escape here

%%include \doinum{number}for the DOI number in the header
%%include \volnum{number} for the volume number in the header
%%include \year{yyyy} for  year of publication in the header
%%include \pgrange{num--num} page range of article in the header
%%include \artcitid{num} for the article citation id
%%include \lp to print last page of the article
%%include \setcounter{page}{pagenum} for the exact starting page of the article

\doinum{12.3456/s78910-011-012-3}
\artcitid{\#\#\#\#}
\volnum{000}
\year{0000}
\pgrange{1--}
\setcounter{page}{1}
\lp{1}

\section{Introduction}
In 1889, George Ellery Hale has built his own spectrohelioscope to make a map of the
sun in particular wavelength (Hale and Ellerman, 1960). Since then, several observatories
around the world have built slit based spectroheliographs which have entrance and exit slits.
The exit slit decides the pass-band of the spectrohelioscope. Later, many observatories started using the filter based spectral isolators to image the sun, particularly the solar chromosphere. Using these instruments, observatories have made full-disk Ca-K and H$_{\alpha}$ images recorded in the photographic plates (Srivastava, Ambastha and Bhatnagar, 1991; Zirin, Ligget, and Patterson, 1982). With the advancement in the detector technology, many have started utilizing the electronic based detectors which are more sensitive to the light and obtained more uniform images (Verma et al. 1997;  Denker et al. 1999; Otruba and Potzi, 2003).

The H$_{\alpha}$ data shows the conditions of the chromosphere during the flare events.
This is the best wavelength region to observe the solar flare ribbons from ground based observatories. The observations made in H$_{\alpha}$ wavelength shows the fibrilar structures which delineates the magnetic fields (Woodard and Chae, 1999). The filament eruption, pre-flare brightening, rotation of prominences before the eruption, dynamics of spicules and many more interesting features could be seen in this wavelength. Mass motions inside the filaments and rotations of the prominences are studied in several cases during its eruption (Zirker {\it et~al.} 1998; Lin {\it et~al.} 2003; Labrosse {\it et~al.} (2010); Mackey {\it et~al}. (2010)). With the high-resolution and long hours of observations of the filaments/prominences it is easy to study their properties using H$_{\alpha}$ observations leading up to the eruption. During the flare, at the chromospheric level two ribbons are seen which move away from each other at certain speed (Vemareddy, Maurya, and Ambastha (2012)). The identification of newly brightened 
pixels can be done using H$_{\alpha}$ wavelength. These newly brightened pixels along with the
corresponding locations in the magnetograms can be used to compute the magnetic flux swept-up by the ribbons
as they move apart in the chromosphere (Qiu {\it et~al} (2004)) and can be used to infer the reconnection rate 
inside the reconnecting region. The details of various other science cases that can be studied with the H$_{\alpha}$ data is reported in Ravindra {\it et al.} (2016).

Kodaikanal observatory has a history of making the full-disk solar observations in
white-light, Ca-K and in H$_{\alpha}$ wavelengths. The white-light and Ca-K observations
have started in 1904 and 1905 respectively. The H$_{\alpha}$ observations have started in 1912. Both the Ca-K and H$_{\alpha}$ observations discontinued in 2007 because the production of photographic plates/films has stopped. Though, the white-light observations using photographic plates continued till today (Ravindra, et al. 2013). Apart from Kodaikanal, the solar observations were also carried out at Aryabhatta Research Institute of Observational Sciences (ARIES) and at Udaipur Solar Observatory (USO) using 15-cm objective, starting from 1972.

To continue the long term full-disk observations of the sun in H-alpha, Indian Institute of Astrophysics (IIA) has worked out the main optical design and configuration of the telescope. Based on the design, two identical telescopes and Lyot filters were  manufactured by Nanjing Institute of Astronomical Optics and Technology (NIAOT), National Astronomical Observatories (NAO), Chinese Academy of Sciences (CAS), China.
Both the telescopes were assembled and tested in NIAOT for its performance.
Out of two telescopes, one of the telescopes was installed at the Kodaikanal Observatory in October, 2014 (Ravindra et al. (2016)). The second one was installed initially at Center for Research and Education in Science and Technology (CREST) campus at Hosakote by our IIA team. We have carried out in-house development of the software control and tracking system of the telescope. After completion of successful testing at CREST Hosakote, telescope was taken to Merak for final installation, which is also the proposed site for the 2-m class National Large Solar Telescope (NLST; Hasan, 2012).

In this paper, we report the installation of the second telescope, development of telescope control software, technique used for guiding the telescope and filter unit. We present some of the representative results in the paper and the paper concludeswith the summary of the telescope installation at Merak.

\section{The Telescope}

\begin{figure*}[!h]
\begin{center}
%\centering
\includegraphics[width=\textwidth]{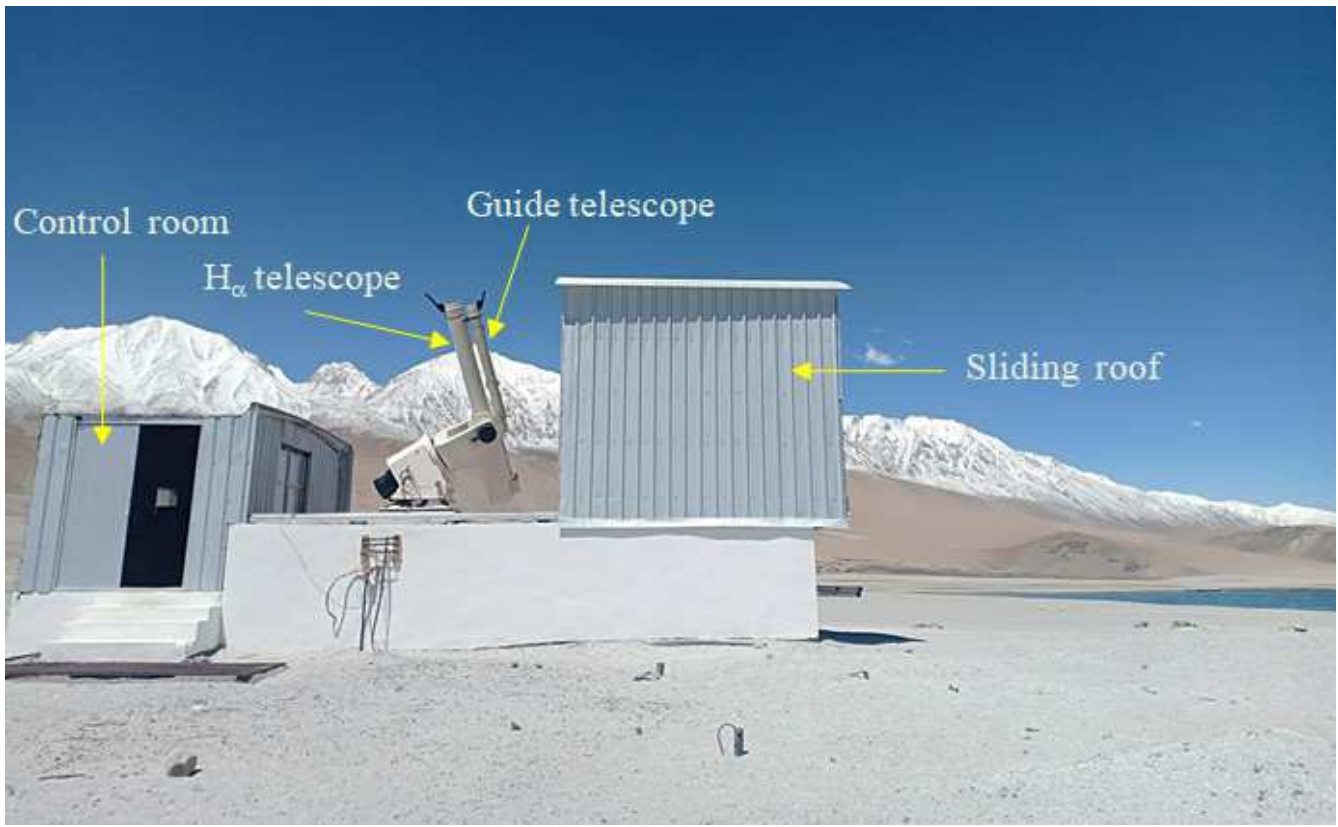} \\
\end{center}
\caption{A picture of the H$_{\alpha}$ telescope making solar observations soon after the installation in August 2017. The telescope is protected by a lightweight sliding roof shown retracted in the image. Control room has the telescope electronics, power supply and computer.}
\end{figure*}

The telescope is installed at Merak village, in the incursion site of Pangong Tso lake.
The Merak village is located about 150~km away in S-E direction from the Leh Town of Ladakh district in Jammu and Kashmir state.  This is also one of the proposed sites for putting up 2-m class National Large Solar Telescope (NLST). The site survey indicates that seeing conditions are conducive for solar observations (Hasan et al., 2010). To take advantage of large water body, the telescope is installed close to the incursion site of Pongong lake.

\subsection{Telescope and Re-imaging optics}

\begin{figure*}[!h]
\begin{center}
\includegraphics[width=\textwidth]{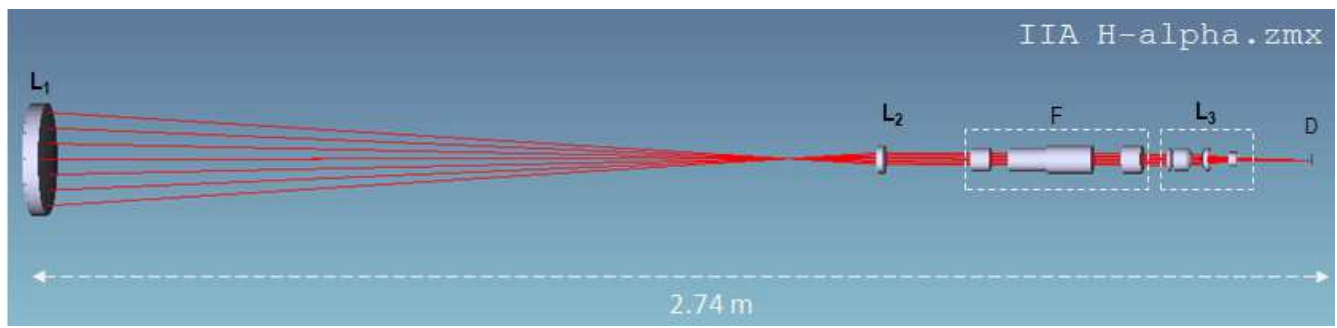} \\
\caption{A schematic of the optical components and the ray diagram. In Figure, L1 is the 20 cm primary lens, L2 is the collimating lens, L3 is the refocusing lens, F is the filter unit and D is the detector position.}
\end{center}
\label{fig:2}
\end{figure*}

Figure~1 shows the picture of the telescope after installation at the incursion site. In this image, the larger diameter tube is the main telescope and the smaller tube is the guide telescope. Figure~2 shows the optical layout of the main telescope designed with Zeemax software. The objective lens, collimating and reimaging optics are all enclosed inside the 3.2~m long tube of the main telescope.   The objective lens is a doublet of diameter 20~cm and focal length of 158~cm. The Lyot filter is kept in between the collimator
and reimaging set-up. This forms a sun's image of size 2.1~cm at the focal plane. A zoom lens which can magnify the sun's image by 2.5 times is kept after the re-imaging lens of the main telescope. The zoom lens is motorized and it can be taken in and out of the beam path whenever required. The main usage of this lens is to make  high-resolution observations of the sun when necessary. A detailed description of the telescope and control system can be found in Ravindra et~al. (2016).

\begin{figure}[!h]
\begin{center}
\includegraphics[width=0.3\textwidth]{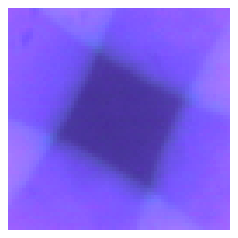} \\
\caption{The image recorded by the guide CCD camera. The non-overlapping darker region in the image is used for measuring the telescope tracking errors.}
\end{center}
\label{fig:3}
\end{figure}

The second smaller tube is a guide telescope. A 10~cm achromatic doublet as an objective lens which is cut into 4 pieces and fixed in a frame which holds each quadrant of the lens with a few millimeter gap in-between them. Each one of these lens produces an image at the focal plane at 4 different corners of a square with atleast 2-sides of each image overlap with each other.  The central portion of the image is dark and has a square shape with sides that are little curved. A CCD camera with 756$\times$562 pixels is used in the focal plane to image the central dark portion. The CCD camera can accommodate only little larger than half  the size of actual image formed that includes the dark portion of the overlapped image. Figure~3 shows the central portion of the image obtained with a split lenses when it is pointed towards the sun and captured by the guide CCD camera. The detected signal is then fed into a Daheng Video controller card for accessing it in computer. The image capture and analysis software is written in the Microsoft--Visual Basic platform.

The guide CCD camera is kept on X-Y linear stage which can be moved with a stepper motor. During the full-disk observations of the sun, the dark portion falls on the center of the tube. But, during the high-resolution observations, the dark portion can fall anywhere on the circle of the tube depending on the region-of-interest. The X-Y translation stages are used to adjust the CCD positions to capture the dark portion onto the CCD camera. 

An algorithm was developed to detect and track the dark portion of the image. The central dark portion of the guide image (figure 3) is converted into bright pixel by inverting the image. The drift in the dark portion of the image is used to find the tracking errors in RA or DEC motors. After the telescope is pointed towards the Sun, the auto guiding is enabled.  The first frame is taken as the reference frame. The subsequent images grabbed then after is used to determine the shift in the telescope position if any with respect to the first frame. The inverted bright region is identified with a suitable threshold.  The pixels in the dark region (inverted region) is multiplied by the corresponding intensity and then the product is summed over all the pixels. This forms the numerator and summed pixel intensity values forms the denominator. The algorithm works on the basis of finding the centroid (Prabhu {\it et~al.} (2018)) of the inverted bright region. 
The shift in the present image centroid is computed with respect to the reference frame centroid. The images are acquired with a cadence of one second to find the shift in centroid. By identifying the shift in the centroid location, shift in RA  and DEC axis are determined.  The calculated shift in both directions are converted into Encoder counts. One pixel shift in auto guide camera corresponds to 14 encoder counts. If the shift is more than two pixels in the auto guide (28 encoder counts) then there will be a feedback sent to the control system to override the tracking and there by correcting the error in the telescope pointing.

A small objective lens, filter and a prism makes a finder telescope which forms a
2.5~cm size image on the ground glass prism. This set-up is used during the initial stage of
telescope alignment. This arrangement is also convenient to show solar images to interested public visiting the observatory.

The main telescope, guider and the finder telescope are all rigidly fixed on common base
that is attached to equatorial fork mount whose polar axis is inclined to 33$^{\circ}$, the latitude of the place. The telescope with pedestal is placed on a reinforced concrete pier with cast iron plate on top of the pier. The height of the pier is about 1~m and the height of the objective lens is about 4~m above the ground when the sun is at the zenith. The mechanical structure of the telescope, direct drive motors and its specifications, break motor units, incremental encoders to get position information of the telescope etc. are all described in detail in Ravindra et~al. (2016). The main technical parameters of the telescope are listed in Table 1.

\begin{table}
\begin{center}
\caption{The specifications of main telescope, guider and finder telescopes.}
%\label(table1}
\begin{tabular}{|l|c|r|}
\hline
Parts & Parameter & values \\
\hline
Main Telescope & Objective size & 20.06~cm \\
               & Focal Length   &  158~cm \\
               & Image size     &  2.69~cm \\
               & Magnified Image size & 6.72~cm \\  
               & CCD pixels           & 2048$\times$2048 \\
               & Pixel size(high-res)   & 0.27$^{\prime\prime}$ \\
               & Pixel size (low-res) & 0.67$^{\prime\prime}$ \\
\hline
Guide Telescope & Objective size & 10~cm \\
                & Focal Length   & 174.3~cm       \\
                & CCD pixels     & 756$\times$562 \\
\hline                
Finder Telescope & Objective size & 3~cm \\
                 & Focal Length & 252.4~cm \\
                 & Image size  & 2.5~cm \\
                                           
\hline
\end{tabular}
\end{center}
\end{table}

\subsection{H$_{\alpha}$ Telescope Control Software}
The H$_{\alpha}$ telescope control software was developed in the Microsoft--Visual Basic platform. The drive units are controlled using National Instruments (NI) cards. The necessary drivers for the NI -- PCI  motion controller card is installed to communicate with the servo motors of the RA and DEC drives.  The telescope tracking system has two incremental encoders in RA and DEC axis respectively.  Optical rotary encoders consist of a light source, rotating code disc, and a light detector. The rotating code disc has graduations with equally spaced areas of dark and bright lines.  As the encoder ring rotates, the light detector registers the ON-OFF pattern. The detector converts the ON-OFF pattern into a digital signal in two channels.  The offset between the two channels determines the direction of the shaft rotation. The technical parameters of the RA and DEC encoders are listed in Table 2.

Each encoder has a reference position. During the initialization, the telescope moves on both axis to find the reference position or home position. From the reference position the location of the sun in the sky is identified using an algorithm. This algorithm was developed using the procedure given in Reda and Andreas (2008), and cross verified with the National Oceanic and Atmospheric Administration (NOAA) solar position calculation $\footnote{\url{https://www.esrl.noaa.gov/gmd/grad/solcalc/NOAA\_Solar\_Calculations\_day.xls}}$.    Once the telescope is pointed towards the Sun, the tracking is enabled and then the telescope starts to track the sun at a speed of 15$^{\prime\prime}$~sec$^{-1}$. 

The telescope is equipped with four IR limit sensors in the four directions. In addition to this, there are four more IR sensors to monitor the limiting positions of the telescope in two  directions. The use of limit sensors prevents any accidental damage to the telescope.  The telescope and the filter units are controlled through Graphical User Interface (GUI)(Figure~4).

\begin{figure*}[!h]
\begin{center}
\includegraphics[width=0.8\textwidth]{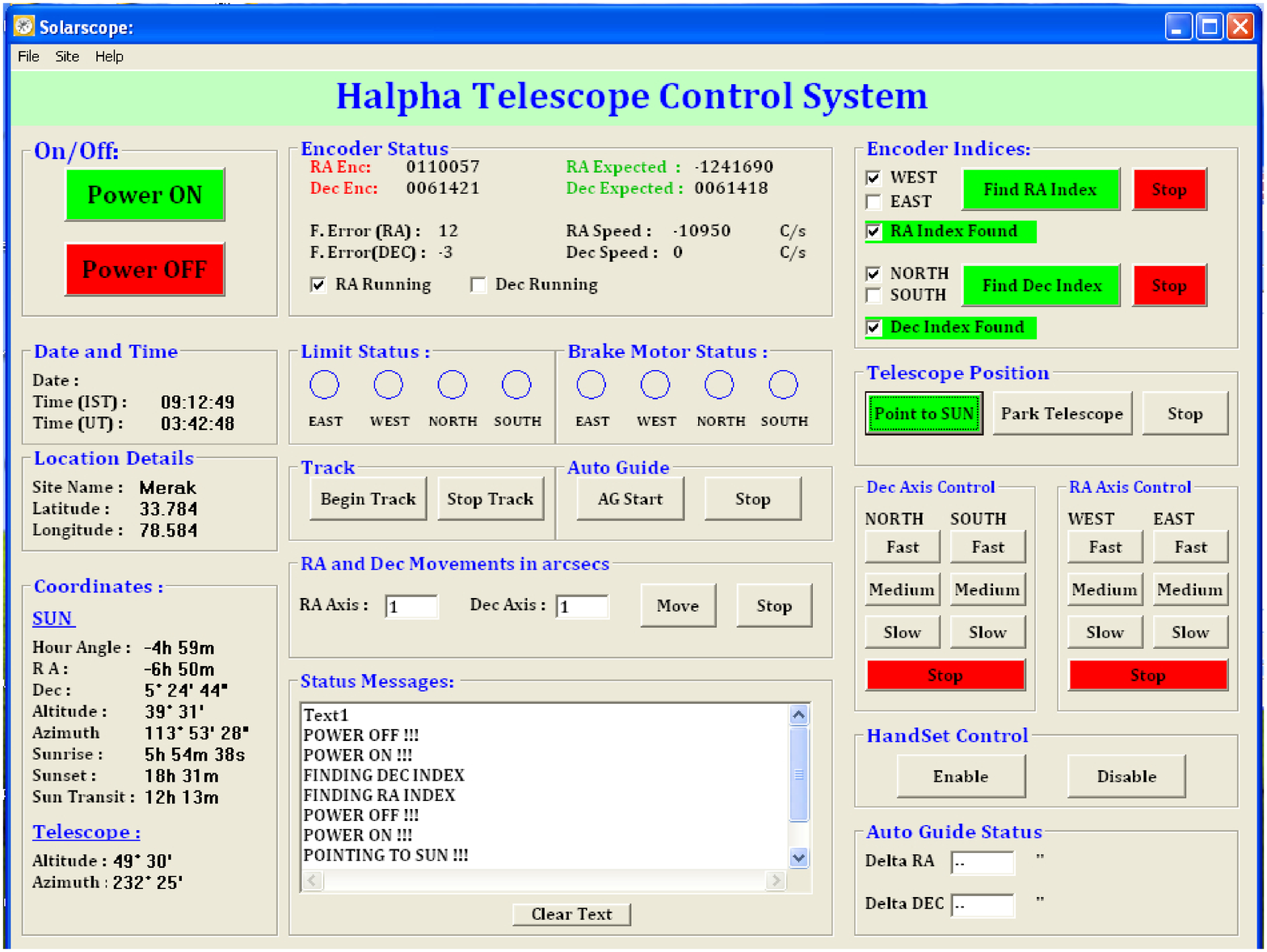} \\
\caption{A screen-shot of graphical user interface developed for telescope control.}
\end{center}
\label{fig:4}
\end{figure*}

\begin{table}
\begin{center}
\caption{The specifications of the encoder unit for the RA and DEC drives at Merak H$_{\alpha}$ telescope.}
\begin{tabular}{|l|r|}
\hline 
Parameters & Value \\
\hline
Type of encoder & Incremental \\
\hline
Encoder power supply  & 5V and 250~mA \\
and current ratings & \\
\hline
Diameter of encoder & 20~cm \\
\hline
Number of lines & 31488 \\
\hline
pitch width & 20$\mu$m \\
\hline
%\label{table2}
\end{tabular}
\end{center}
\end{table}

\subsection{H$_{\alpha}$ Spectral Line Isolator}
Solar atmosphere can be imaged by observing the sun in isolated spectral passband.
There are several ways to isolate the solar spectral line. Interference filters whose 
passbands of $\sim$~1~\AA~or less is widely used to image solar chromosphere.
However, the image quality is not the best due to poor contrast. Some of the observatories use mica based Fabry-Ferot interferometers whose passband could be anywhere between 0.4 - 0.6~\AA~(GONG; Hill, et al. 2009). Some others used Lithium Niobate based or liquid crystal variable retarders (LCVRs) based, voltage tunable filters (Tomczyk, Mathew and Gallagher, 2016; Hagino et al., 2014). Birefringent crystals based Lyot filter (Bethge et al. 2011),  Michelson interferometer based filter used in Michelson Doppler Imager (MDI; Scherrer et al., 1995) and Solar Dynamics Observatory (SDO; Scherrer et al., 2012) etc. Compared to Fabry-Perot based filters, birefringent filters have a large field-of-view but the throughput is less. 

The design used in Lyot filter of H$_{\alpha}$ telescope installed at Merak has 7 stages 
of Calcite crystals. The filter bandwidth is of 0.5~\AA~and the free spectral range (FSR) is about 52~\AA. The whole assembly is motorized to tune the wavelength positions. The filter
can be tuned as far as 4~\AA~away from the line center with a spectral position accuracy of
10~m\AA. The complete filter assembly was immersed in a silicone oil chamber to ensure a uniform temperature distribution. Since the refractive index of the crystals and the silicone oil almost match with each other, the use of silicone oil reduces the internal multiple reflections between the crystals at the interface.
The temperature is controlled by two stage heating around different layers of the filter unit. The system can work with the ambient temperature ranging from --20$^{\circ}$~C to 35$^{\circ}$~C. A 30~\AA~prefilter is kept in front of the Lyot filter to isolate the H$_{\alpha}$ spectral line and reject all the other side bands. The entire Lyot filter assembly along with the prefilter was kept in a collimated beam to avoid any shift in the passband due to beam convergence. The operating principle of Lyot filter unit is explained in Stix (2002).  More details about the Lyot filter used in H$_{\alpha}$ telescope
at the Kodaikanal Observatory which is identical to the one used at Merak H$_{\alpha}$
can be found in Ravindra et~al. (2016). A few technical details about the Lyot filter is listed in Table 3.

\begin{table}[!h]
\begin{center}
\caption{The specifications of H$_{\alpha}$ Lyot filter existing at Merak telescope.}
\begin{tabular}{|l|r|}
\hline
Parameters & Value \\
\hline
Central Wavelength & 6562.82~\AA~\\
\hline
Passband           & 0.5~\AA~\\
\hline
Free Spectral Range & 51.2~\AA~\\
\hline
Pre-filter passband & 30~\AA~\\
\hline
Scanning Range     & 4~\AA~on both sides \\
                   & from the central wavelength \\
\hline
Scanning step size & 10~m\AA~\\
\hline
Temperature and stability & 42$^{\circ}$C and 0.01$^{\circ}$C \\
\hline
Number of stages of crystal & 7 \\
\hline
Acceptance angle & 2$^{\circ}$ \\
\hline
%\label{table3}
\end{tabular}
\end{center}
\end{table}

\subsection{The Detector Unit}
The Charge Coupled Device (CCD) is an Apogee make (Alta series), used to record the image produced by the telescope with H$_{\alpha}$ filter. The camera is 2048$\times$2048 pixels. 
The pixel size (7.5~${\mu}$m) corresponds to 0.27$^{\prime\prime}$ in high-resolution mode.
The typical full-well capacity of the camera is 40000 electrons for this
series. The CCD is front illuminated with quantum efficiency little larger than
30\% at the H$_{\alpha}$ observing wavelength. The digital resolution of the camera
is 16-bits and the CCD image can be read out with a clock speed ranging from 1 -- 8 MHz. It has
a interline transfer sensor with an exposure time varying from 1~ms to several minutes.
The CCD can be cooled to 50$^{\circ}$~C below the ambient. Usually, the day time temperature at Merak is less than 15$^{\circ}$~C and hence we expect the CCD to operate around --30$^{\circ}$~C or less. At this temperature the dark current is about 0.01~e$^{-}$~pixel$^{-1}$~s$^{-1}$. The CCD is interfaced with the data acquisition
system through USB~2.0 port.

To get an image with full-well capacity, an exposure time of 400--500~ms was given. 
Though it is possible to take the images with 10~ms or better exposure time, the readout time is large between 0.5--4 sec. Hence, the data cannot used for speckle imaging to get 
diffraction limited images. Also, at this exposure time the number counts in the CCD pixels is very less.  
In future upgrades we will use high speed cameras with exposure time of few milli second to carry out such observations. The exposure time varies from morning to evening. Normally, the images were taken at a cadence of 10~sec and in case of any events occurring on the sun, the images acquired at much higher cadence. The whole software to get an image was written in Python software which not only controls the main telescope CCD camera, but also controls the Lyot filter stepper motors to move the spectral window on the H$_{\alpha}$ line profile.

\section{Telescope Installation and Operation}
The telescope was installed in the month of August, 2017 at the Merak village near the
Pangong Tso lake. The width of the lake is about 2-km and the length is
about 130~km. The maximum depth of the water is about 100~m. The large water body  near
the incursion site makes the seeing better which in turn improves the image quality of the telescope.

The telescope was installed on August 30, 2017 at the Merak site. When the telescope was
installed, two large sunspot groups were present on the sun, one in the Northern
(AR NOAA 12674) and another one in Southern hemisphere (AR NOAA 12673).  Since, the  detector 
area is small, it is not possible to cover the full-disk of the sun in any mode of operation.  
When these active regions appeared on the sun,
we obtained their images  at a cadence of 10~sec in high-resolution
mode. The dark exposures are taken once a day after the observations finished,
just before closing the telescope. This is simply because the camera does not
have any shutter. After closing the telescope shutter, no light enters the
telescope and hence it becomes easy for obtaining the dark exposures.  Flat-field images
are taken once or twice a day during the observations. These images are taken
by moving the telescope in all the possible directions by small amount. The daily observations
are made in H$_{\alpha}$ line center at a cadence of 10~sec. These images
are sent to IIA, Bengaluru for further calibration and scientific analysis.

\begin{figure*}[!h]
\begin{center}
\includegraphics[width=0.45\textwidth]{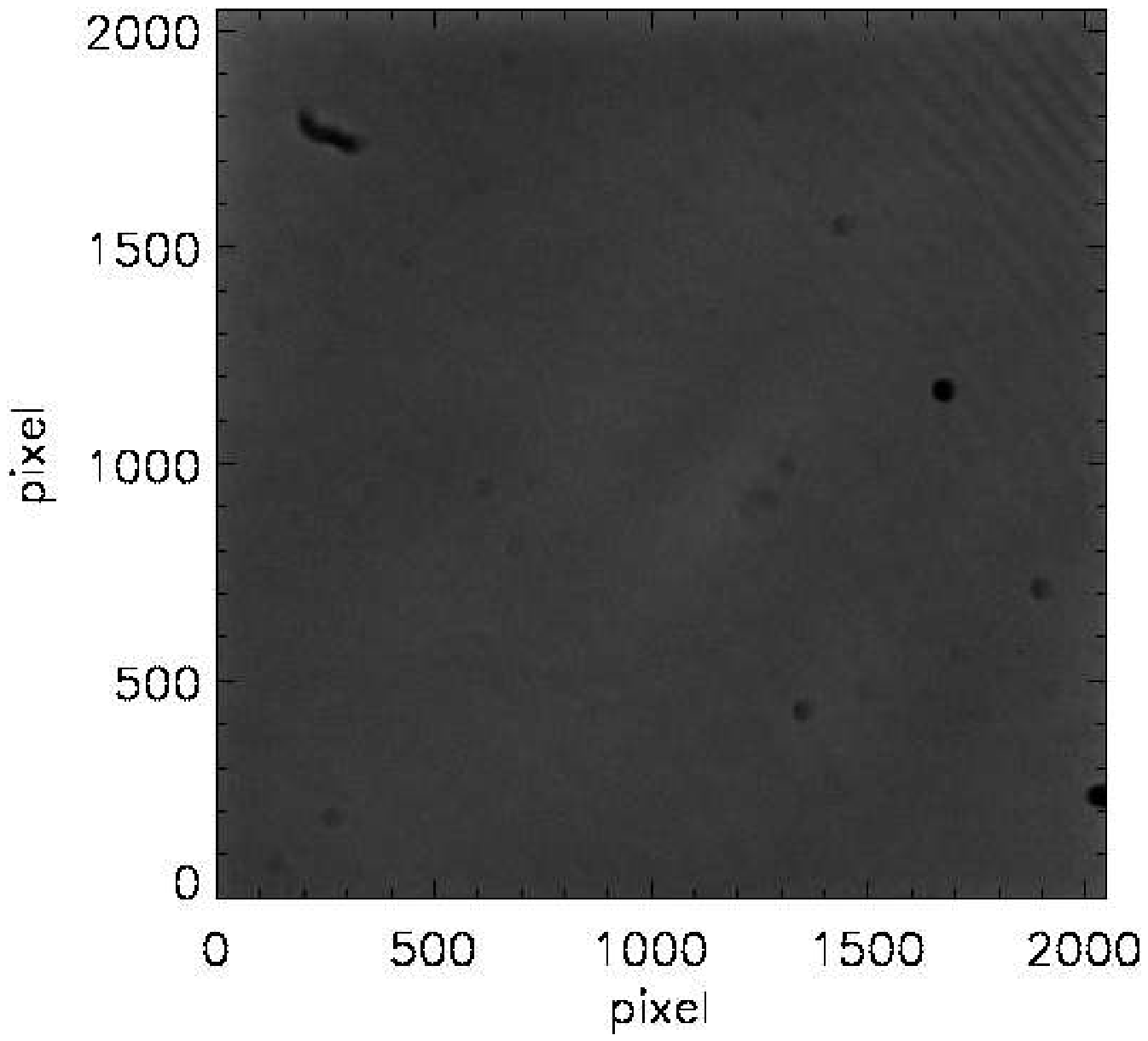}\includegraphics[width=0.45\textwidth]{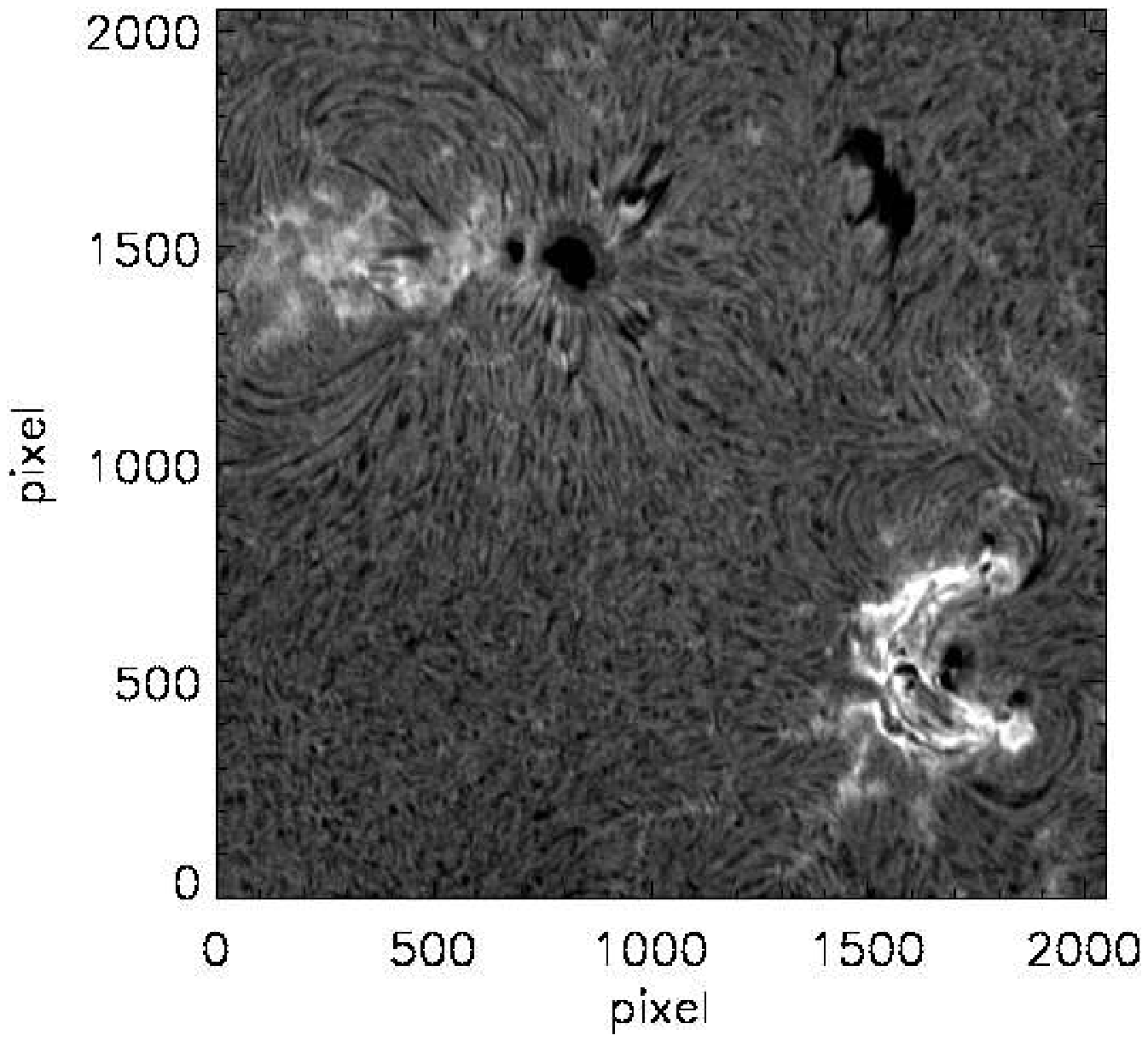}
\end{center}
\caption{Left: The left side image is master flat obtained after using the Chae (2003) method. Right: The dark subtracted and flat fielded H$_{\alpha}$ image obtained on September 05, 2017 at 05:46~UT. The image field-of-view is 9.2$^{\prime}$.}
\label{fig:5}
\end{figure*}

The H$_{\alpha}$ data needs to be processed before it is made useful for
the scientific analysis. The images are flat-fielded using the master flat.
The following procedure is used to obtain the master flat. We obtained the shifted
images while taking the observations. The telescope is shifted by about 10 pixels
in random directions and about 10 to 12 images are obtained
this way. The dark exposures are also taken and averaged to construct a master dark frame. Then
the shifted images and master dark are used to make a master flat using the method
described in Chae et~al. (2004). A sample master-flat image is shown in
Figure~5 (left). The dust particles and other non-uniformity in the data
can be seen in the flat. The right side image shows the flat-fielded image.

\subsection{Image co-alignment}
The obtained images are co-aligned with the first image in the time series. This has been done by finding the correlation after shifting the images in
X and Y directions.  The images are then co-aligned with the sub-pixel accuracy with the first image in the time series. Later, we co-align these images
with the white-light image taken by Solar Dynamics Observatory. While aligning the image we made sure that the timings of the images in
H$_{\alpha}$ and white-light are the same. These co-aligned images are useful for studying multi-wavelength nature of the transient events.

\section{Preliminary data and Results}

\begin{figure*}[!h]
\begin{center}
\includegraphics[width=0.45\textwidth]{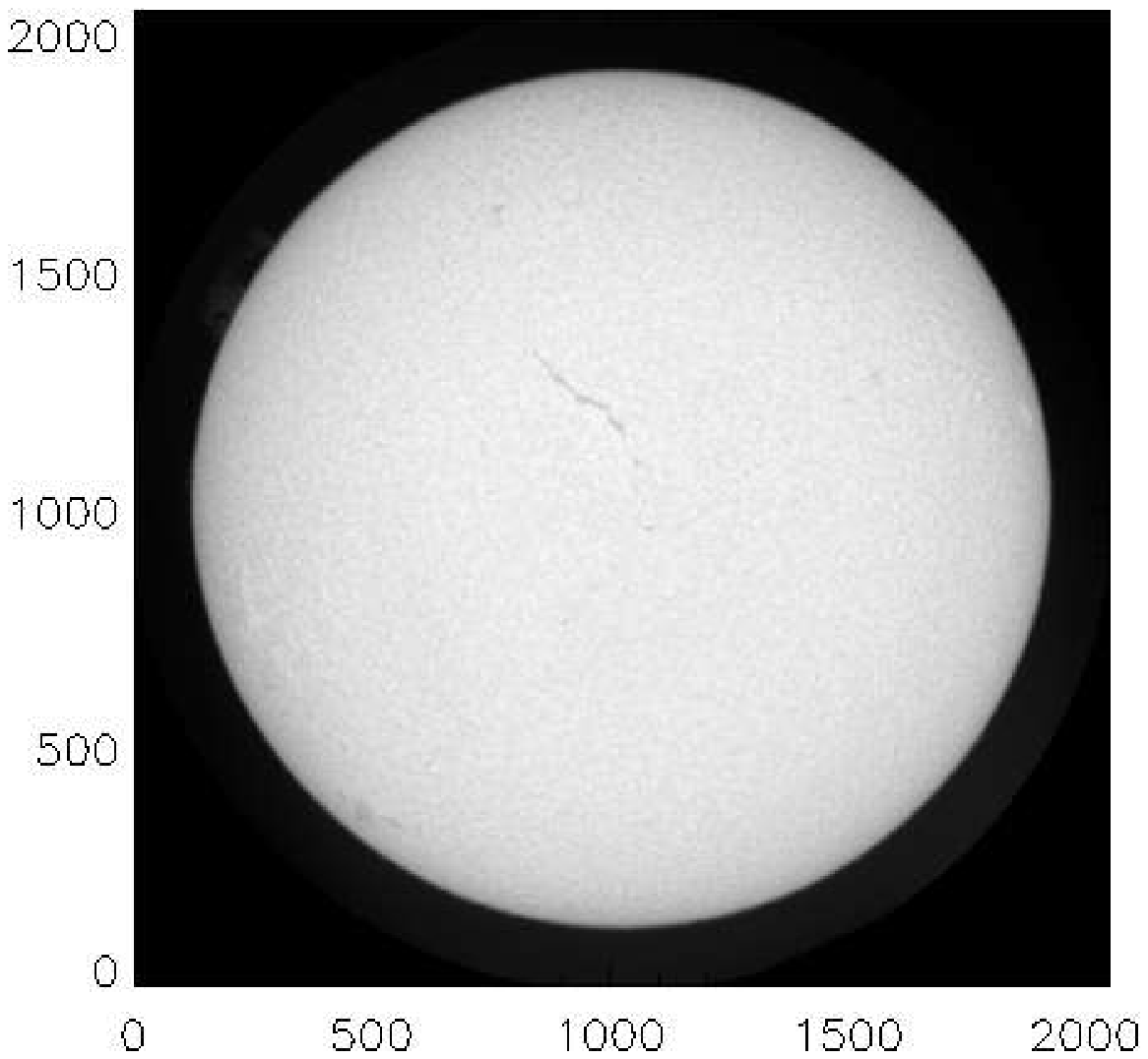}\includegraphics[width=0.46\textwidth]{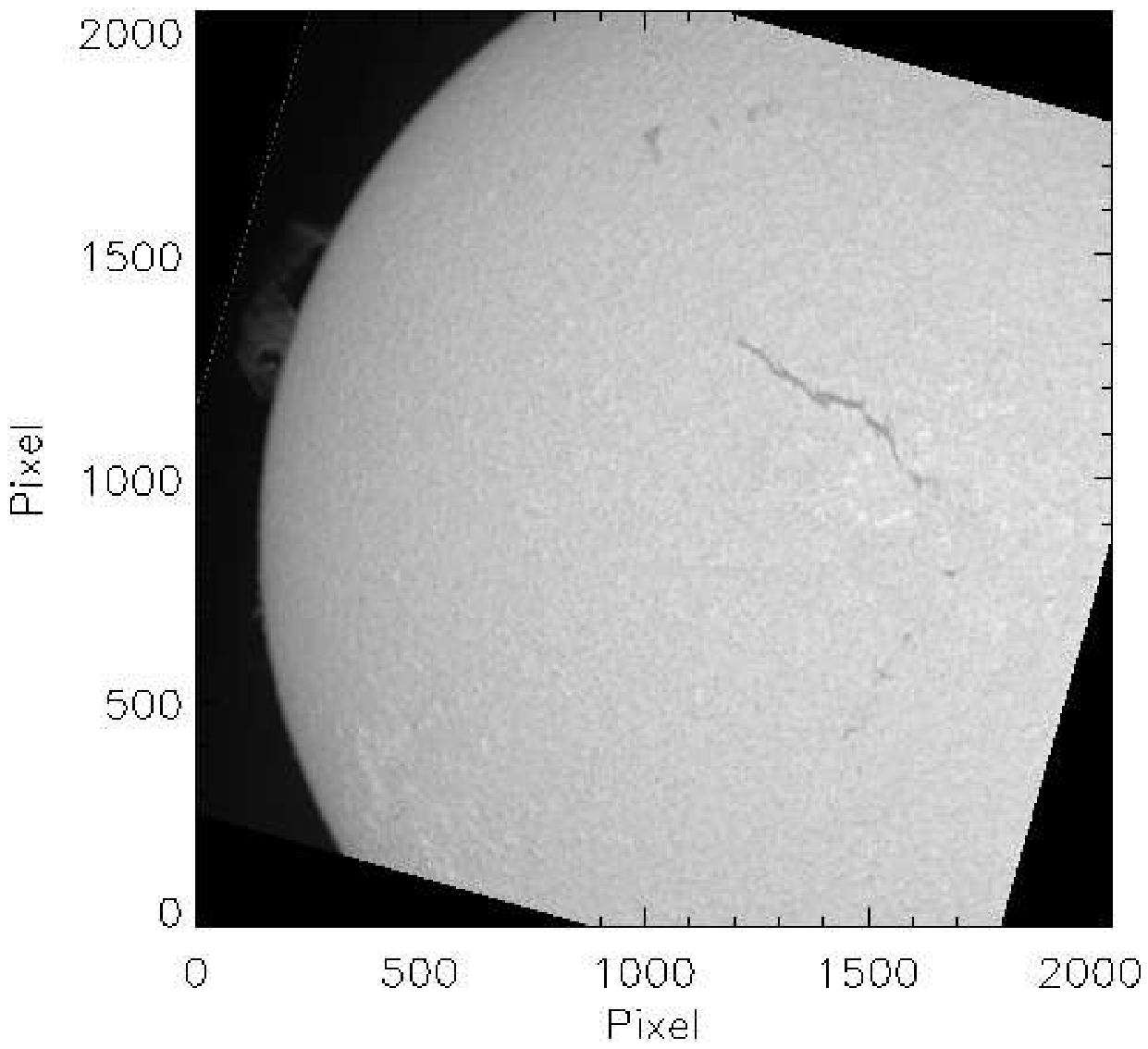}
\end{center}
\caption{Left: The GONG H$_{\alpha}$ image obtained from Udaipur station on 19 October 2017. Right: The H$_{\alpha}$ image obtained from Merak site on the same day and close in time to GONG image. The pixel resolution of GONG image is about 1$^{\prime\prime}$ and the Merak image is 0.68$^{\prime\prime}$.}
\label{fig:6}
\end{figure*}

Global Oscillation Network Group (GONG; Harvey, 1988) has updated its optics in 2010 to provide the H$_{\alpha}$ images to the solar community (Harvey, et~al. 2011). GONG uses 7-cm objective lens and its pixel resolution is 1$^{\prime\prime}$. It obtains the H$_{\alpha}$ images in all the 6 stations.  Figure~6 shows the comparison of GONG (Udaipur) and Merak H$_{\alpha}$ images. Obviously, Merak image has smaller FOV (with pixel resolution of 0.68$^{\prime\prime}$) because of the small format CCD camera. In both images it is easy to identify the filaments and plage regions. In the Merak image, the large prominence on the limb is clearly seen while barely visible in the GONG image. Apart from the large filament structure (which is lying in the East-West direction), there is one more filament below it lying in the North-South direction. Again, this is clearly seen in the Merak H$_{\alpha}$ image but not unambiguous in the GONG image. There is one more quiet sun filament which is laying horizontally in the Northern hemisphere at high latitude, but not seen clearly in the GONG image. For this event, we did not have any observations from Kodaikanal observatory to make any meaningful comparison with its counterpart at Merak. 

\begin{figure*}[!h]
\begin{center}
\includegraphics[width=0.5\textwidth]{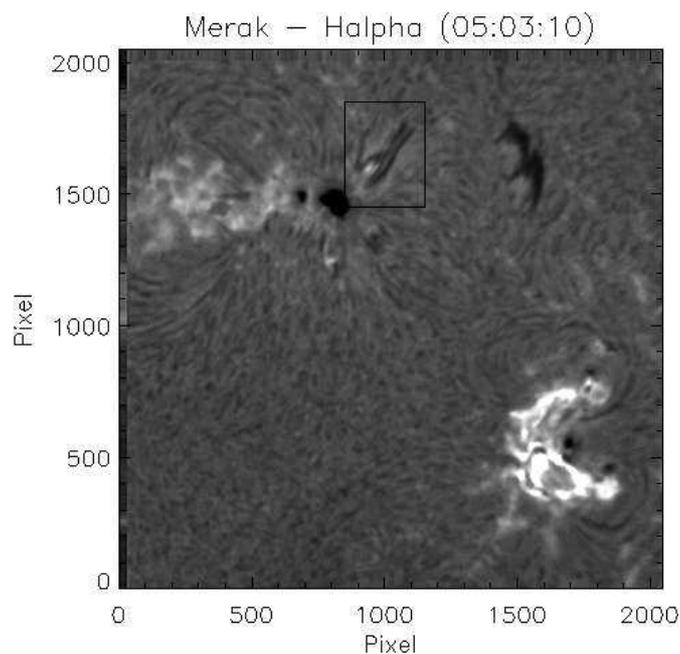} \\
\end{center}
\caption{An image of the two active regions, filament and fibrils surrounding the sunspots
are shown. A surge is shown by a boxed region near the active region NOAA 12674. The field-of-view of the image is 9.2$^{\prime}$.}
\label{fig:7}
\end{figure*}

The active regions NOAA 12673 and 12674 appeared in the Southern and Northern hemisphere
respectively, during the declining phase of the solar cycle 24. On September 05, 2017 both active regions have crossed the central meridian passage. On September 06, 2017 the active region 12673 produced X9.3 class flare at 11:53~UT. On September 05, it produced several M and C class flares. The AR 12674 was little quiet compared to other one. However, it has produced several surges during its lifetime. Figure~7 shows the location of surge occurred near the active region NOAA 12674. The surge location is shown in the boxed region.

\begin{figure*}[!h]
\begin{center}
\includegraphics[width=0.9\textwidth]{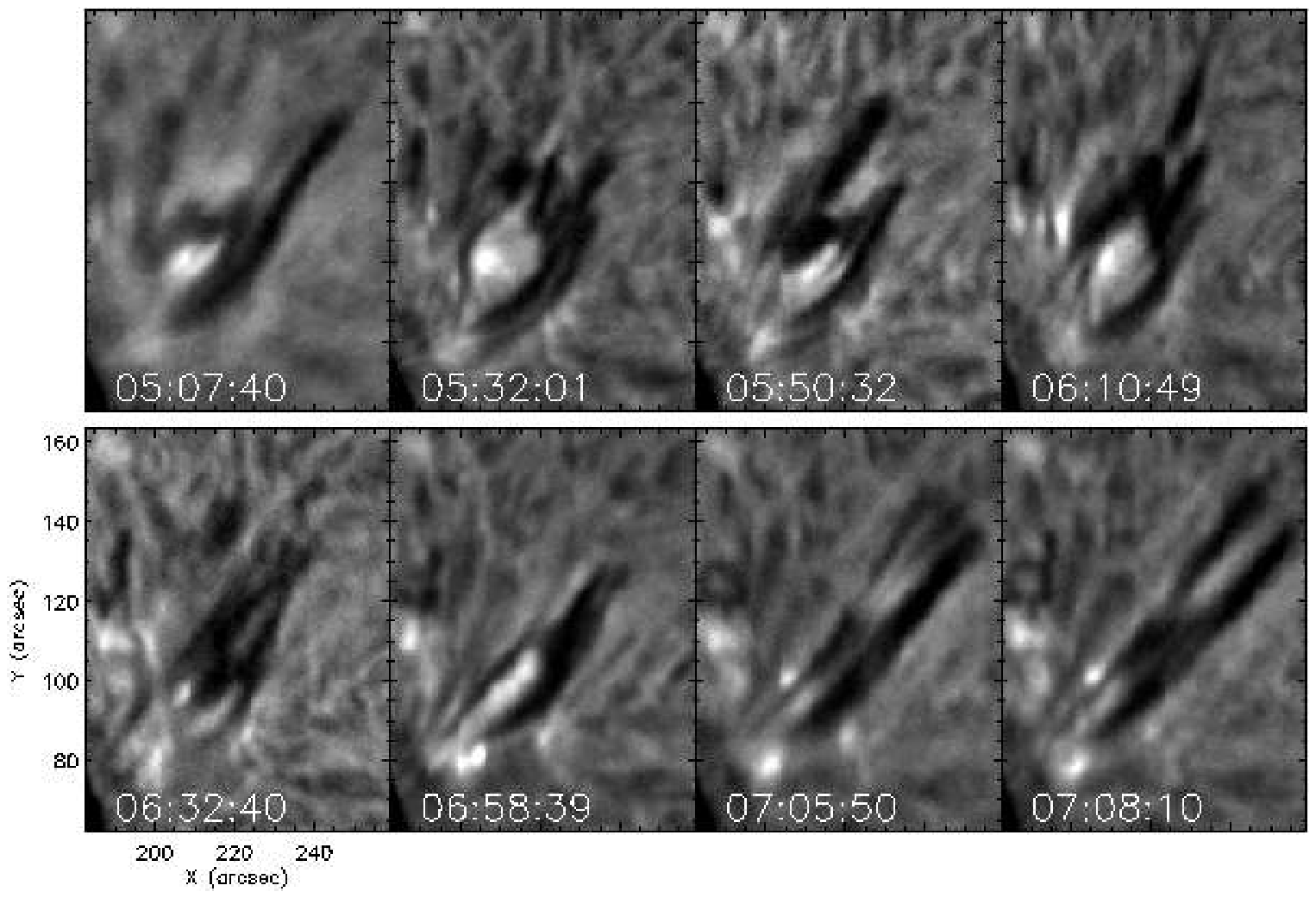} \\
\end{center}
\caption{A time sequence of images showing the evolution of H$_{\alpha}$ surge.}
\label{fig:8}
\end{figure*}

Figure~8 shows a time sequence of H$_{\alpha}$ images extracted from the boxed region. The surge eruption initiated at 03:10~UT. During the eruption phase (05:32~UT) it appeared as funnel shaped and then it shrunk in width (06:32~UT). It elongated at 06:58~UT and erupted at 07:08~UT. Overall it took about 4~hour to erupt. The surge had bright ejecta surrounded by the dark ejecta in the funnel shape. A close look at the surge in the initial phase indicates that it is a filament and ejected as a surge.

\begin{figure*}[!h]
\begin{center}
\includegraphics[width=0.5\textwidth]{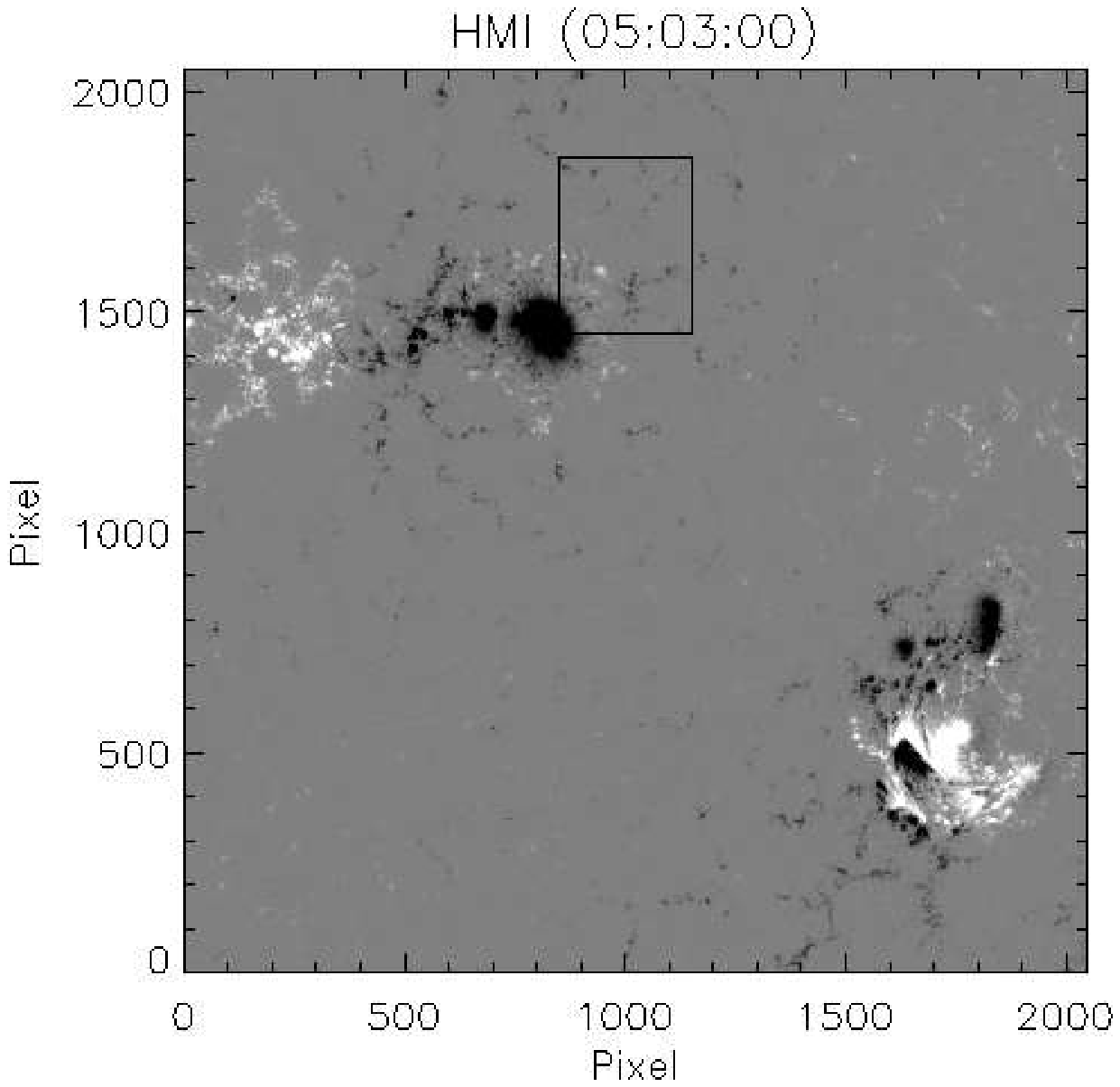} \\
\end{center}
\caption{The image of magnetogram corresponding to the same field-of-view as H$_{\alpha}$ image shown in Figure~7. The boxed region shows the location of jet in magnetogram.}
\label{fig:9}
\end{figure*}

\begin{figure*}[!h]
\begin{center}
\includegraphics[width=0.9\textwidth]{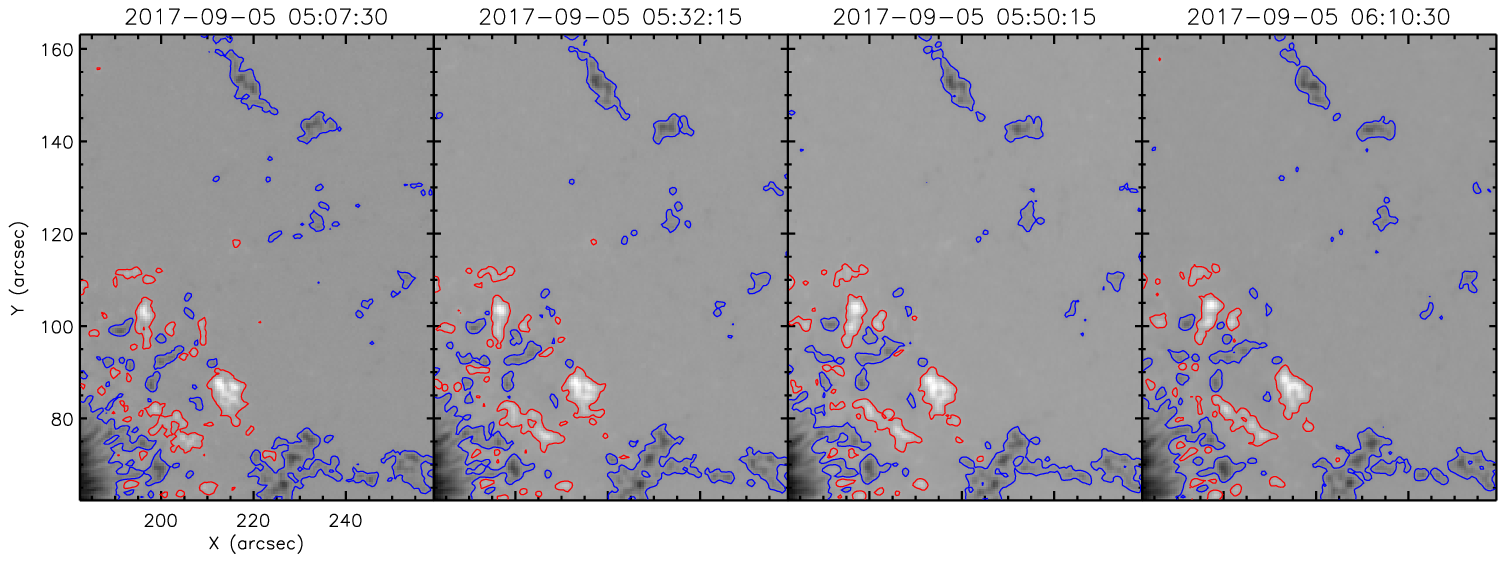} \\
\end{center}
\caption{A time sequence of images showing the evolution of magnetic fields in the location of jet. The Field-of-view is same as Figure~8.}
\label{fig:10}
\end{figure*}

Figure~9 shows a magnetogram of the same FOV as Figure~7. The magnetogram is obtained from HMI instrument onboard the Solar Dynamics Observatory (SDO; Pesnell, Thompson,  Chamberlin, 2012). The boxed region shows the surge location. Clearly the leading sunspot in AR 12674 is of negative polarity and the surrounding region is dominated by moving magnetic features with mixed polarity (MMFs: Harvey and Harvey, 1973). Figure~10 shows the evolution of the magnetic fields in the surge location. Clearly, there is a collision of southern polarity region with the large northern polarity region followed by cancellation.  The surge gets erupted during the disappearance of the southern polarity region.  A more detailed result from this event will be presented in a future publication.

\begin{figure*}[!h]
\begin{center}
\includegraphics[width=0.9\textwidth]{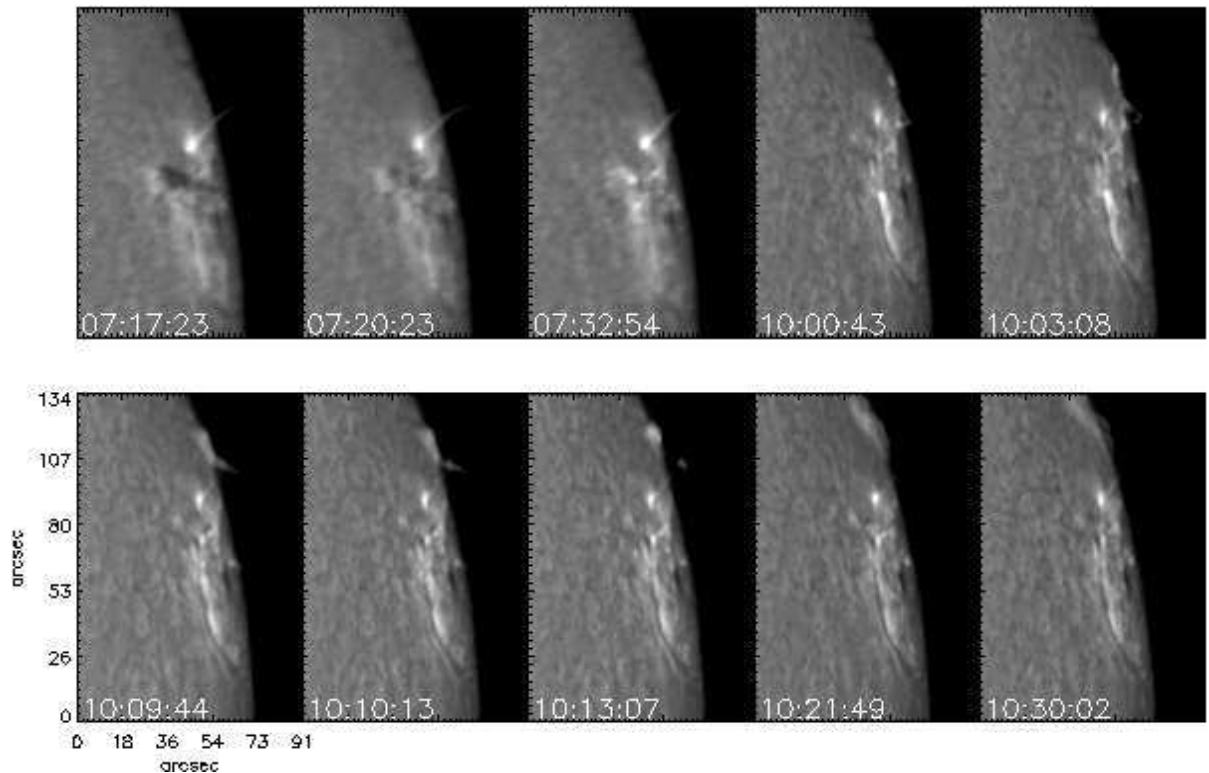} \\
\end{center}
\caption{A time sequence of surge occurred in the limb near the AR 12673 observed on September 9, 2017.}
\label{fig:11}
\end{figure*}

Figure~11 depicts the surge event near the active region 12673 when it moved towards the limb on September 9, 2017 at 07:17~UT. The surge erupted in about half an hours time and several surges frequently observed in that region. During our observation period, the region ejected 7 surges. Many of them were failed ones, they all fell back to the nearby prominence region.

\section{Summary}
Indian Institute of Astrophysics (IIA) has planned and designed the H$_{\alpha}$ telescope. Two identical telescopes were fabricated at NIAOT, china and tested for its performance.
The NIAOT has made the 2 Lyot filters whose passbands are 0.4 and 0.5~\AA. A team from NIAOT has installed one of the telescopes at the Kodaikanal Solar Observatory in 2014 to continue the observations in H$_{\alpha}$ wavelength which was initiated in 1912 at Kodaikanal.
The Lyot filter installed at Kodaikanal has a passband of 0.4~\AA. The second one is installed at the Merak site by IIA team, near the Lake shore of Pangong Tso, where the IIA is planning to setup a large 2-m class telescope dedicated for solar observations. The H$_{\alpha}$ telescope is installed at Merak site in August 30, 2017. Since, then it is making the observations of the sun whenever the sky is clear. The passband of the Lyot filter installed at Merak is 0.5~\AA. The chromospheric images obtained from the telescope shows fine features in H$_{\alpha}$ wavelength and showed several surges in many active regions. The H$_{\alpha}$ telescope installed at Merak site provides another addition to the existing solar facilities at Kodaikanal.

The Lyot filter has the facility to tune it to any position on the H$_{\alpha}$ line profile. A stepper motor attached to the filter unit can be tuned to any point on the line profile with an accuracy of 10~m\AA. The tuning takes about 1~s to move the central passband from one position to the other on the line profile. This property of the filter can be used to reconstruct the line profile and also can be used to find the Doppler shift in the line profile position. Using this, it is possible to make dopplergrams at the chromospheric heights (Dhara, Ravindra and Banyal 2016; Chae, Park and Park 2006; Joshi et~al. 2013). In future, we will produce such dopplergrams and make it available for scientific use.

At present, a smaller size CCD camera is used. This covers only 9.2$^{\prime}$ FOV. In future, we would like to replace it with a large format CCD to cover the entire 32$^{\prime}$ FOV.  The Merak H$_{\alpha}$ telescope is a new addition to the existing facilities at Indian subcontinent. This trans Himalayan region largely remains unaffected by regular monsoon and thus enable useful and uninterrupted observations of the sun.

\section*{Acknowledgments}
We thank Prof. Jadev Singh for initiating the H-alpha telescope program in IIA. The telescope and H$_{\alpha}$ filter were fabricated and tested at Nanjing Institute for Astronomical Optics (NIAOT), CAS China.
We thank the telescope team at NIAOT for their help and support. Special thanks to
Prof. Weijun Mao for his constant encouragement and support during the
development of the control software for the telescope at IIA.
We also would like to thank the team members at IIA for their help during
the installation of the telescope:  Nagaraj, B,
Manoharan, J., Tundup Stanzin, Tsewang Punchok, Tundup Thinless,
Stanzin Norbo, Kunzhang Paljor, Konchok Nima, Ramaswamy, M. V, Ramachandra Reddy R, Srinivasa,
Parthiban, D., Vinay Kumar Gond, Pandiyarajan, B., Anand, Periyanayagam, Francis, Ramesh, Sreeramulu Nayak, P. R., Manjunath, Rajkumar and Mahendran A. Without their help it would not
have been possible to install the telescope during those
difficult conditions. Thanks also to the director of IIA,
Dr. Sreekumar and Prof. Rangarajan for their constant support and encouragement
during the testing phase of the telescope at Hosakote and installation of
the telescope at Merak. We Thank the referees for their valuable comments which improved the manuscript.

\end{document}